\documentclass[twocolumn,showpacs,nofootinbib,tightenlines,nobibnotes,
aps, amssymb,amsmath,prl]{revtex4}
\usepackage{graphicx}
\usepackage{epsfig}

\newcommand{\la}{\langle}
\newcommand{\ra}{\rangle}
\newcommand{\ua}{\uparrow}
\newcommand{\da}{\downarrow}
\newcommand{\nn}{\nonumber}

\newcommand{\be}{\begin{eqnarray}}
\newcommand{\ee}{\end{eqnarray}}
\newcommand{\cdi}{c^\dagger_{i\sigma}}
\newcommand{\cdj}{c^\dagger_{j\sigma}}
\newcommand{\ci}{c_{i\sigma}}
\newcommand{\cj}{c_{j\sigma}}

\begin{document}
\preprint{APS/123-QED}
\title{
Resonant Pair Exchange and Percolation in the Disordered Hubbard Model}

\author{Dariush Heidarian$^1$, Mustansir Barma$^1$ and Nandini Trivedi$^{1,2}$}
\affiliation{%
$^1$ Department of Theoretical Physics,
Tata Institute of Fundamental Research,
Homi Bhabha Road, Colaba, Mumbai 400005, India\\
$^2$ Department of Physics, The Ohio State University,
190 W. Woodruff Avenue, Physics Research Building, Columbus, OH 43210, USA.
}%

\date{\today}

\begin{abstract}
We show that the effect of disorder on a Mott-Hubbard insulator of interacting electrons
produces a quantum phase transition from an
antiferromagnetic to a paramagnetic phase
governed by a percolation of the singly occupied weakly disordered sites. 
Near the transition, we propose that a new type of defect, formed by the resonant exchange between a spin singlet and a doubly occupied site with an
attractive disorder potential, plays an important role. 
These resonant pair exchange defects reduce the staggered magnetization but enhance the coupling of the two spins 
and produce characteristic signatures in the temperature dependent 
specific heat and the non-Curie spin susceptibility, 
at temperatures on the order the hopping $t$,  
higher than the typical 
exchange scale $J$.  
\end{abstract}

\vskip2pc

\maketitle


Recent theoretical and experimental studies show that the interplay of
random disorder with correlations in electronic systems can produce
striking effects.  Such studies have largely focused on the metal
insulator transition\cite{lee1985,krav94, hanein98}, 
where the problem still remains open from the
theoretical perspective.  Also of great interest is the question of
magnetism in strongly correlated disordered systems. Besides the
destruction of magnetic long range order with increasing disorder, 
we show in this Letter that disorder induces new types of magnetic coupling, absent
in the ordered system.

We address these questions by studying the effect of disorder on the
magnetic properties of the Hubbard model.  Besides its importance as a
theoretical paradigm, this model is also relevant for the high $T_c$
superconductors, where mobile carriers are produced by stoichiometric
doping that is also a source of quenched disorder.  At half filling,
in the absence of disorder, the model describes a Mott insulator with
a spin at every site, and antiferromagnetic (AF) coupling of
order $J\sim t^2/U$ between spins at
adjacent sites. 
We show that in the presence of disorder,
it is also
possible to have a novel type of magnetic coupling that we call {\em
resonant pair exchange} (RPE), which operates between particular pairs
of sites. This interaction is the outcome of the resonance to order
$t$ between two configurations of spins on adjacent sites: (a) a spin
singlet formed by single spins on adjacent sites and (b) a
non-magnetic doublon formed by a doubly occupied and empty pair due to
a binary potential close to ($U/2$ and $-U/2$) at two nearest neighbor
sites.  This sort of `defect pair' occurs with a finite probability
in a disordered system, and has several important characteristics: (i)
Similar to two-level systems in glasses, these defects produce a
characteristic maximum in the specific heat.  
(ii) The staggered spin susceptibility is suppressed because of the 
mixing of the singlet configurations with a non-magnetic configuration. 
However, interestingly, the non-Curie behavior persists to temperatures 
$T\sim t$, which is much higher than the kinetic exchange scale
$J\sim t^2/U<t$. (iii) The resonant tunneling produces high kinetic
energy on the bond connecting the two sites. This provides a source of
noise in RPE defects that should be trackable in conductance noise
experiments\cite{popovic}.

These RPE defects are most active near the antiferromagnetic (AF) to 
paramagnetic (PM)
transition driven by increasing disorder. 
Strong disorder generates
two types of sites: non-magnetic unoccupied or doubly occupied sites,
and magnetic sites with a single spin\cite{singh}. 
We find that a percolation-based description
then becomes possible, as electron hopping results in coupling between
neighbouring magnetic sites; increasing the disorder leads to a
decreasing number of magnetic sites, and eventually to a transition
marking the loss of long range AF order.  We supplement
this with other calculational approaches, including an inhomogeneous
Hartree Fock (HF) calculation whose results are found to agree remarkably 
well with quantum Monte Carlo simulations\cite{vajk, sandvik}.


{\it Model:} We start with the repulsive Hubbard model on a two-dimensional square lattice
with site disorder:
\be
H&=&-t\sum_{\la ij\ra,\sigma}(\cdi\cj+\cdj\ci)+ U\sum_in_{i\ua
}n_{i\da}\nn\\
&+&\sum_{i\sigma}(V_i-\mu)n_{i\sigma}
\label{eq:H}
\ee
where $c_{i\sigma}^\dagger (c_{i\sigma})$ is the electron creation 
(annihilation) operator with spin $\sigma$ on site $i$ and 
$n_{i\sigma}=\cdi\ci$ is the corresponding density 
operator. 
The first term describes hopping of electrons between nearest neighbors with 
amplitude $t$. 
The second term is the repulsive on-site interaction for doubly occupied
sites with strength $U$, $\mu$ is the chemical
potential that determines the filling and 
$V_i$ is a random potential
chosen from a uniform distribution between $-V$ and $V$.

{\it The Large $U$ limit}: 
In the atomic limit $t = 0$, the competition between repulsive
interactions and disorder produces site-dependent integral occupancies $n_i$
in the ground state.  Sites with $V_i > U/2$ are unoccupied;
and those with $V_i < -U/2$ are doubly occupied; neither of these two types 
of sites has a free spin.
On the other hand, sites with $|V_i| < U/2$ have $n_i = 1$ with a free spin
residing on each such site.  Thus the fraction of randomly placed 
singly-occupied 
(magnetic) sites is $x = U/2V$, while the remaining sites are
nonmagnetic. 

In the atomic limit, the spin degeneracy of the ground state is
$2^{xN}$, where 
$N$ is the total number of sites.
The effect of turning on a small hopping amplitude $t$ is
to lift this degeneracy and produce an AF coupling of magnitude
$J_{ij}$ between nearest-neighbor (nn)
singly-occupied sites
$i$ and $j$, by the well-known mechanism of kinetic exchange.  
To second order in $t$, the coupling $J_0=4t^2/U$ in the pure system is modified 
by the disorder difference $\delta V=V_i - V_j$ on the two sites to
\begin{equation}
J_{ij} = \frac{J_0}{1-(\delta V)^2/U^2}\  \cdot
\label{eq:Jij}
\end{equation}
The effective leading-order Hamiltonian is then
${\cal H} = \sum_{\langle ij \rangle} J_{ij} S_i \cdot S_j$
where $J_{ij}$ is nonzero only if both sites are magnetic.

{\it Resonant Pair Exchange}: 
It is evident from the above analysis that there will be some rare regions
where the disorder at a pair of neighbouring magnetic 
sites $V_i$ and $V_j$ is such that
the conditions $|V_i - V_j - U| < t \ll U$ and $|V_{i/j}|\simeq U/2 $ hold.  
In that case, the denominator in
Eq.~\ref{eq:Jij} becomes very large, and the perturbative expression is no
longer valid. In fact, in this regime the electron hopping couples these
pairs of sites to {\it first} order in $t$, and we show that 
it induces a new type of coupling, which we call resonant pair exchange (RPE).
The RPE process differs qualitatively from normal kinetic exchange,
and has important consequences for 
the thermodynamic and transport properties of the system.

Consider states with 2 electrons on a pair of sites characterized by
disorder parameters $V_1$, $V_2$, when the hopping $t=0$.  Of the
total of 6 states, there are 3 singlet $(S=0)$ states
$|1,1\ra_s, ~~|2,0\ra,~~|0,2\ra$.
and 3 triplet
$(S=1)$ states
$|\ua,\ua\ra,~~|\da,\da\ra,~~|1,1\ra_t$.
Since the Hamiltonian conserves total spin $S$, we
examine each subspace separately. Of the three singlet states,
two states $|2,0\rangle$ and $|0,2\rangle$ involve unequal charges at
each of the two sites, whereas one state $|1,1\rangle_S = 1/\sqrt{2}
(|\uparrow,\downarrow\rangle - |\downarrow,\uparrow\rangle)$ involves
one electron on each site.  We are interested in the case when one of
the two unequal-charge states is nearly degenerate with $|1,1\rangle_S$.  For
specificity, let us take $V_1 = -V_2 = U/2$.  Then 5 of the six states
(the 3 triplet states and 2 singlet states $|1,1\rangle_S$ and
$|0,2\rangle$) are degenerate with energy $-U$,
while $|2,0\rangle$ has energy $U$.
\begin{figure}[tbp]
\includegraphics[width=7.5cm]{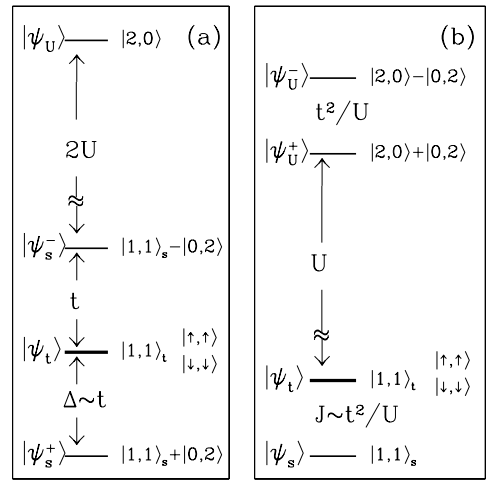}
\caption{
Eigenstates for two sites. (a) Resonant pair exchange: 
With $V_1=-V_2=U/2$; the difference in energy between the RPE 
singlet ground state and triplet excited state is $\Delta\sim t$. 
(b) Kinetic exchange: With $V_1=V_2=0$; the energy difference between
the magnetic singlet ground state and the triplet excited state is 
$\sim J\sim t^2/U$.}
\label{f:psi}
\end{figure}
The primary effect of nonzero but small hopping is to mix the two
degenerate singlets, and produce the eigenstates 
\be
|\psi_s^\pm\rangle=\frac{1}{\sqrt{2}}\biggl(|1,1\rangle_S \pm
|0,2\rangle\biggr),
\label{three}
\ee
with eigenvalues $-U-\sqrt2t$ and $-U+\sqrt2t$ respectively. 
The resulting pattern of energy levels is as shown in
Fig.~\ref{f:psi}a.
\begin{figure}
\epsfxsize=.8\hsize
\centerline{\epsfbox{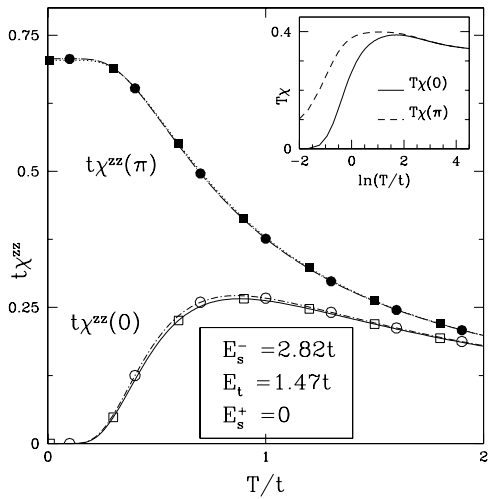}}
\caption{
$\chi(\pi)t$ and $\chi(0)t$ versus $T/t$ for different
$U/t$ for two sites at half filling and with $V_1=U/2$ and $V_2=-U/2$.
The square symbol is for $U=8t$ and the circle symbol for $U=16t$.
When scaled by $t$, the curves for different T and $U$ collapse to one curve. 
The inset depicts $T\chi$
versus $\ln(T/t)$, showing that $\chi$ has Curie behavior at large $T$.
\label{f:X2}}
\end{figure}

In the case $V_1 = V_2 = 0$,
the states $|2,0\rangle$ and $|0,2\rangle$ have a
high energy $U$, while the hopping lowers the energy of
$|1,1\rangle_S$ with respect to the triplet state by an energy of
order $t^2/U$ (Fig.~\ref{f:psi}.b).  This is the mechanism of kinetic
exchange, which produces AF correlations in the dimer
ground state that persist up to $T\sim t^2/U$.  With resonant pair exchange, by contrast, the
AF correlations in the ground state are reduced (as the
non-magnetic state $|0,2\rangle$ is mixed in); but these correlations persist to
much higher $T\sim t$ than in the case of kinetic
exchange. As a consequence, 
both the uniform susceptibility $\chi(q=0)$ and the
ordering susceptibility $\chi(q = \pi)$ show 
deviations from Curie (free moment) behaviour for 
$T<t$, signaling the onset of AF
correlations due to resonant pair exchange, 
as shown in Fig.~\ref{f:X2}.
Similar magnetic effects are expected whenever
the conditions $|U + V_i - V_j| \leq t$ and $|V_{i/j}|\simeq U/2$ 
are satisfied for
neighboring pairs of sites $\langle ij \rangle$.  The fraction of
such pairs in the disordered Hubbard model is 
of the order of $(t/2V)^2$. As the contribution
to energy lowering from each pair is $t$, the
overall contribution to the ground state energy is of order $t^3/V^2$, 
which is of higher order in $t$ than from
the majority of pairs, which are coupled by normal kinetic
exchange.

\begin{figure}
\epsfxsize=7cm \centerline{\epsfbox{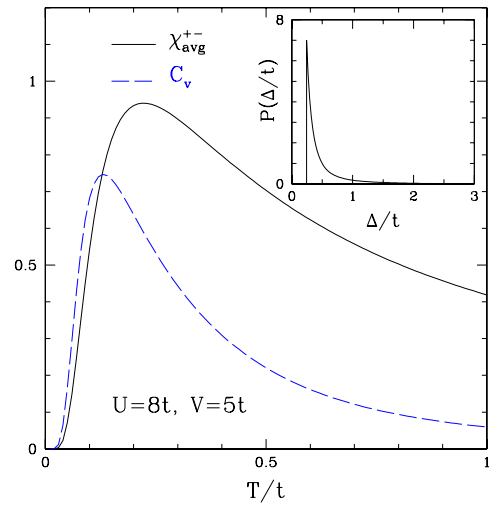}}
\caption{The averaged susceptibility $\chi^{+-}_{avg}$ 
and specific heat versus the
temperature. Both quantities show a peak at $T\sim J$ where $J\sim t^2/U$ is the exchange coupling.
The resonating pair exchange defects contribute at higher temperatures
$T\sim t$. The distribution of the
energy level splittings is shown in the inset and shows that it is
dominated by small splittings $\sim J$.
\label{f:specific}}
\end{figure}

{\it Specific Heat and Spin Susceptibility}:
Since sites coupled by resonant pair exchange are relatively rare  and
have a very different level structure (see Fig.~\ref{f:psi}) 
from the majority of pairs, they act as localized centers and give rise
to a distinctive signature in the specific heat $C_v$ and $\chi_{avg}$,
much as two level centers do in glasses. For an arbitrary pair, 
the energy splitting $\Delta$ between the singlet ground state 
and the triplet states is
\be
\Delta=
\sqrt{2t^2+\left(\frac{U-\delta V}{2}\right)^2}-\frac{U-\delta V}{2} 
\label{eq:Delta}
\ee
where $\delta V=|V_1-V_2|$. Here we have assumed the state with energy $U+\delta V$
decouples from other two levels. If $V_1$ and  $V_2$ are chosen from a uniform
distribution between $-V$ to $V$ and 
we regard each pair as isolated from the others, 
the probability distribution for the splitting $\Delta$ is given by 
\be
P(\Delta)=\frac{1}{2V}\left[\frac{\Delta}{V}\left(\frac
{4t^4}{\Delta^4}-1\right)+
\frac{2V-U}{V}\left(\frac{2t^2}{\Delta^2}+1\right)\right].
\label{eq:distribution}
\ee
The total specific heat from such pairs is
$C_v=\int_{\Delta_{min}}^{\Delta_{max}}d\Delta
P(\Delta)c_v(\Delta)$
where $\Delta_{min}$ and $\Delta_{max}$ can be obtained by 
substituting $0$ and $U$ for $\delta V$ in Eq.~\ref{eq:Delta}, 
since the integration is only over 
singly occupied sites. 
Noting that the average energy for this two level system 
is $ E(\Delta)=-\Delta\exp(\Delta/T)/\bigl[\exp(\Delta/T)+3\bigr]$,
the corresponding specific heat is 
$c_v(\Delta)\equiv{\partial E}/{\partial T}$, shown in
Fig.~\ref{f:specific}.

We see that at $T\geq t$ most of the contribution to the 
total specific heat
is from RPE sites, due to their large splitting. In this regime $C_v$
varies as $1/T^2$. On lowering $T$, $C_v$ has a peak at 
$T=2t^2/\lambda U$ with $\lambda \simeq 2.85$. 
Most of the peak weight comes from pairs with small 
splitting ($\Delta\propto t^2/U$). 
In the low-$T$ regime, $C_v$ decays exponentially as $T\rightarrow0$; 
this form of the decay found within the pair approximation would
change if the system supports extended (spin wave like) states. 

The above arguments hold also for the averaged susceptibility. 
For the two level system mentioned above 
spin susceptibility is $\chi^{+-}(\Delta,T)=2/T[\exp(\Delta/T)+3]$. 
The susceptibility averaged over pairs shows Curie behavior ($\propto1/T$) 
at high $T$ and a peak at $T\approx 2t^2/U$ (see
Fig.~\ref{f:specific}); below this temperature, 
triplet states make a very small contribution to the susceptibility. 
As $T$ tends to zero $\chi^{+-}_{avg}$ falls as $\exp(-\Delta/T)$.
Therefore the high temperature ($T\geq t$) behavior of the specific heat 
and susceptibility is determined by the RPE sites while the low
temperature behavior is governed by pairs of sites with small energy splittings. 

{\it {Percolation of Magnetic Sites}}: 
Recall that in the atomic limit, a fraction $x=U/2V$ of sites is magnetic 
with $n_i=1$, while the remaining sites with $n_i = 0$ or $2$, 
are nonmagnetic. 
Since only nearest neighbour magnetic sites are coupled if $t$ is small,
the system maps to a dilute magnet\cite{stinchcombe}.


A necessary condition for antiferromagnetic long range order (AFLRO) 
is that there be an infinite connected
cluster of singly occupied sites, i.e. $x>x_c$ where $x_c$ is the site
percolation threshold for a given lattice geometry\cite{footnote1}. 
Thus in the limit
$t/U \ll 1$, we can rigorously rule out LRO for $x<x_c$.  Further, in
this limit, the model reduces to a dilute Heisenberg
AF with random couplings. The question is then whether
there is AFLRO for $x>x_c$. Since $J_{ij}$ in Eq. 2 is bounded below by
$J_0$, and a dilute AF with uniform coupling between
magnetic sites on a square lattice seems to show AFLRO\cite{vajk,
sandvik, Chernyshev}, we expect AFLRO in the disordered Hubbard model 
with $t/U\ll 1$, provided that $V<U/2x_c$.
The occurrence of a small number of resonant pairs in the infinite 
cluster would result
in a slight loss of AF order in the ground state. 
However, if the temperature is much larger
than $t^2/U$ but still of the order of $t$, only the resonant pairs
would make a non-Curie contribution to magnetic properties.


\begin{figure}
\epsfxsize=7cm \centerline{\epsfbox{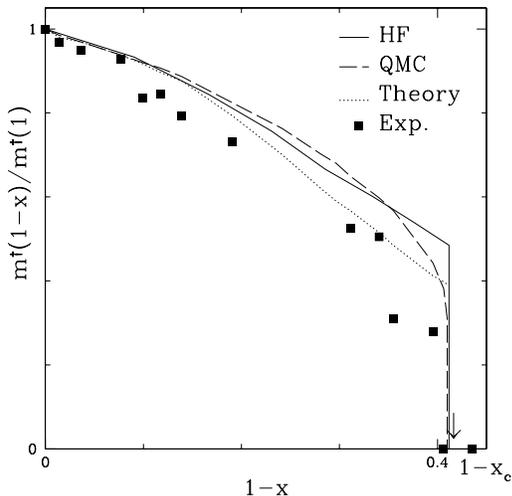}}
\caption{
Solid line is AF order parameter $m^\dagger$ vs
$1-x=1-U/2V$ using HF approximation for a 2D square lattice of the
model mentioned in Eq.~\ref{eq:H} the system size is $28\times28$,
$U=4$ and $T=0$.
The dashed and dotted lines are
staggered magnetization vs dilution in the 2D quantum Heisenberg
model with spin-1/2 using quantum Monte Carlo\cite{vajk,sandvik} and 
spin wave theory\cite{Chernyshev}. 
The square symbols are experimental data points using neutron
scattering\cite{vajk}. The percolation threshold is $1-x_c\simeq 0.41$.}
\label{f:perc}
\end{figure}

{\em Results of the Hartree-Fock approximation}:
The HF approximation, allowing for inhomogeneity in spin and densities 
on all the sites, sheds light on both magnetic and  conducting properties of
the disordered Hubbard model. Earlier work using this approximation focused on 3D\cite{dasgupta} and the effects of moving 
away from half filling. Here we focus on the case of half filling in 2D. 
We find that adding site disorder leads to closing of 
the Mott gap in the density of states (DOS), while
AFLRO persist upto the percolation threshold of magnetic sites,
$2V_{cr}\simeq U/0.59$ (Fig.~\ref{f:perc}).
At an intermediate regime of disorder and interaction the system has
metallic behavior\cite{DHNT}.
Paramagnetic sites start appearing only when $V> U/2$.
For disorder strength $V>V_{cr}$ the system
breaks into clusters of AF sites with {\em no} long range order,
and displays a glassy behavior; the final state of
self-consistent iterations depends on the initial inputs of the
variational parameters of the trial HF Hamiltonian.
With increasing disorder the size of AF clusters shrinks further, and at
the limit of very large disorder the system is a paramagnetic Anderson
insulator.

Figure~\ref{f:perc} shows a good consistency of
the staggered magnetization obtained within the HF approximation
with results of other theoretical and experimental studies of the 
diluted Heisenberg model. 
The HF approximation captures rather subtle effects as well. In
particular the next nn coupling, coming from a fourth order expansion in
$t$  can compete with the nn coupling to produce occasional mismatches in
the alignments of particular spins/clusters --- and these are reproduced
by HF treatment. 
Further for a honeycomb lattice at half filling the AFLRO vanishes at a
disorder strength, predicted by the percolation picture. 

{\it Conclusion}:
There are two significant results in this paper: 
Firstly, we have identified a new type of disorder induced defect that involves
pairs of binary potentials and produces a magnetic state with a reduced staggered magnetization but with an enhanced coupling.   
Secondly, we have shown that the disordered Hubbard model at large $U$ can be mapped to a disordered
diluted AF spin-1/2 quantum Heisenberg model. This allows us to connect the existence of AFLRO with a percolating infinite cluster of magnetic sites with $|V_i|<U/2$.
We find a remarkable consistency between the 
percolation picture and HF results even for intermediate $U$.

\end{document}